\documentclass[a4paper]{article}
\usepackage[utf8]{inputenc}   
\usepackage{amsmath,amsfonts,amssymb,amstext,amsthm,amscd, mathtools}         

\usepackage{cite}
\numberwithin{equation}{section}
\usepackage[hidelinks]{hyperref}
\usepackage{subfig}           
\usepackage{latexsym}         
\usepackage{graphicx}         
\usepackage{float}            
\usepackage{here}             
\usepackage{multicol}         
\usepackage{grffile}          
\usepackage{hyperref}         
\usepackage{wrapfig}
\usepackage[usenames]{color}

\usepackage{fancyhdr}
\usepackage[paperwidth=8.5in, paperheight=11.0in, top=1.0in, bottom=1.0in, left=1.0in, right=1.0in]{geometry}
\begin{document}


\title{Hyperscaling violating Schrödinger black holes in Einstein-Maxwell-scalar theory }
\author{\small
{Alfredo Herrera-Aguilar\footnote{aherrera@ifuap.buap.mx}$\ ^{,1}$, 
Joannis E. Paschalis\footnote{paschalis@physics.auth.gr}$\ ^{,2}$,
Carlos Eduardo Romero-Figueroa\footnote{cromero@ifuap.buap.mx}$\ ^{,1}$
}\\
\small{\textit{$^1$Instituto de Física, Benem\'erita Universidad Aut\'onoma de Puebla,  C.P. 72570 Puebla, México.}}\\
\small{\textit{$^2$Theoretical Physics Department, Aristotle University of Thessaloniki, 54124 Thessaloníki, Greece.}}
}
\date{\small{\today}}

\maketitle
\begin{abstract}
We present a novel family of asymptotically Schrödinger hyperscaling violating black holes with a generic dynamical critical exponent  $ z $ and an arbitrary number of spacetime dimensions. This black hole family represents a solution within the Einstein-Maxwell-scalar setup with a self-interaction scalar potential where the Maxwell field is coupled to the scalar field. 
Through an analysis of the curvature invariants it is observed that this configuration is asymptotically regular for different ranges of the hyperscaling violating exponent $\theta$. 
Furthermore, the above mentioned solution constitutes a gravitational candidate for describing field theories with hyperscaling violating Schrödinger symmetry at finite temperature within the framework of the Gravity/Condensed Matter Theory correspondence.
\end{abstract}

\section{Introduction}
The holographic correspondence is a conjecture that allows us to study strongly coupled quantum field theories using gravity systems defined in a higher dimension. In the holographic context, a $ D $ dimensional field theory is mapped to a $ D + 1 $ dimensional gravitational theory, where the field theory lives on the conformal boundary of spacetime. The initial interest in holography arose with conformal field theories dual to Anti de Sitter spacetime (AdS/CFT)\cite{maldacena1999large}. However, to describe condensed matter systems in the vicinity of a critical point, these holographic techniques need to be extended beyond the relativistic domain, since in the proximity of a critical point many systems exhibit anisotropic dynamical scalings. A prototype example of such critical points is a Lifshitz fixed point where an anisotropic scaling in the time and space directions $(t,x)\rightarrow (\lambda^z t,\lambda x)$ takes place. Such a scaling is characterized by a critical dynamical exponent $z$, where theories  with $ z \not = 1 $ are invariant under non-relativistic transformations \cite{kakru, taylor2008non}. Systems exhibiting this type of symmetry have a great possibility of having gravitational duals that can be realized as appropriate deformations of AdS, being Lifshitz and Schrödinger spacetimes the simplest ones.\\

Lifshitz geometries are dual to scale invariant field theories. Although they are not conformally invariant, these spacetimes have a dynamical critical exponent $ z \not = 1 $ that breaks Lorentz invariance and an extra dimension or holographic coordinate $r$. Originally introduced in \cite{kakru}, they are understood as gravitational duals to non-relativistic quantum field theories at zero temperature. The metric of a Lifshitz spacetime takes the form
\begin{equation}
\label{metrica1}
    ds^2=-\frac{dt^2}{r^{2z}}+\frac{1}{r^2}\bigg[dr^2+dx^idx_i\bigg], \quad \quad i=1,2,3 \ldots D;
\end{equation}\\
 where $ z = 1 $ corresponds to the relativistic AdS spacetime, which is generated by the Einstein-Hilbert action with a negative cosmological constant. The metric (\ref{metrica1}) is a generalization of AdS with an extra parameter and is widely used as an \textit{ansatz} to model condensed matter systems at zero temperature like quantum critical points \cite{kakru,taylor2008non,hartnoll, dong2012aspects, taylor2016lifshitz}. 
 
 Other spacetimes that can be used on the gravity side to reproduce the anisotropic scaling between time and space are Schrödinger geometries \cite{duval,golub}. The group of isometries of these geometries is called the Schrödinger group $ Sch_D (z) $ \cite {guica, son2008toward}. This non-relativistic group can be realized geometrically in light cone coordinates by first deforming AdS and thus reducing the symmetries to those of the Schrödinger group \cite {son2008toward}. Furthermore  $ Sch_D (z) $ is presented as the isometry group of the following $D+3$ dimensional geometry \cite{taylor2016lifshitz,son2008toward}
\begin{equation}
    ds^2=\frac{-b^2(dx^+)^2}{r^{2z}}+\frac{1}{r^2}\bigg[dr^2+dx^jdx_j+2dx^+dx^-\bigg],\quad j=1,2,3...D,
    \label{metrica3}
\end{equation}
where the constant $ b $ is arbitrary and corresponds to a deformation of the conformal theory\footnote{Since the case $ b = 0 $ corresponds to the AdS metric \cite{taylor2016lifshitz}.}. One finds that the generators of the Schrödinger algebra correspond
to the following isometries of the metric (\ref{metrica3}) 
\begin{equation}
\label{cono}
 \begin{split}
H: \quad x^+\rightarrow x^{'+}=x^+ + a,\quad
M: \quad x^-\rightarrow x^{'-}=x^- + a;
\end{split}
\end{equation}
 \begin{equation}
 \label{traslaciones}
\begin{split}
P^{i}: \quad x^i \rightarrow x^{'i}=x^i + a^i, \quad
L^{i j}: \quad x^i \rightarrow x^{'i}=L^i_j x^j\\
\end{split}
\end{equation}
\begin{align}
 \label{Dz} D_z: \quad x^i \rightarrow x^{'i}=\lambda x^i, \quad x^+\rightarrow x^{'+}&=\lambda^z x^+, & x^-\rightarrow x^{'-}=\lambda^{2-z} x^-, \quad \quad &r \rightarrow r^{'}=\lambda r.
 \end{align}
\begin{equation}
C^i: \quad x^i\rightarrow x^{'i}=x^i - v^ix^+, \quad x^-\rightarrow x^{'-}=x^- +v^ix^i-\frac{1}{2}v^2x^+
\end{equation}
Naturally, while breaking the Lorentzian symmetry, Lifshitz and Schrödinger backgrounds are not expected solutions to pure gravity environments. For instance, zero-temperature Schrödinger geometries are solutions within theories of gravity coupled to a massive vector field \cite{taylor2016lifshitz} or within topologically massive gravity models where the usual Einstein-Hilbert action is replaced by a Chern-Simons gravitational term \cite{guica}. On the other hand, geometric aspects of Schrödinger spacetimes in Poincare and global coordinates are studied in \cite{blau2009geometry}. Moreover, one of the main works related to Schrödinger backgrounds discusses the possibility of these geometries to correspond to zero-temperature gravitational duals of a strongly coupled unitary fermion system with a Schrödinger symmetry group \cite{son2008toward}.\\

To holographically study field theories at finite temperature, the holographic dictionary demands the gravitational dual solutions to be black hole configurations \cite {hartnoll, zaanen}. However, the occurrence of black holes in non-relativistic spacetimes requires certain enhancement of the theory through the inclusion of additional matter fields. Namely, the first asymptotically Lifshitz back hole was reported in \cite{taylor2008non} in the context of Einstein-Maxwell-Dilaton theory. Since then, many static asymptotically Lifshitz black holes have been found within several gravitational theories. However, intrinsically Schrödinger black holes with arbitrary dynamical exponent $z$ have not been reported so far. The only published solution corresponds to an $AdS_5$ planar black hole in special light-cone coordinates with $z=2$ \cite{kim2011properties}. In this work the thermodynamic and transport properties of the obtained Schrödinger spacetime is studied through a boost. Subsequently, a further generalization of this work to an arbitrary number of spacetime dimensions with $z=2$ was reported in \cite{sadeghi2013thermodynamics} where it was noticed that adding a scalar field with a self-interaction potential in the action renders a solution that supports the hyperscaling violating symmetry. Likewise, more elaborated black hole solutions have been obtained in \cite{adams2008hot}, where it was realized that the five dimensional finite temperature Schrödinger spacetime with $z=2$ can be obtained from an $AdS_5$ x $S^5$ solution of type IIB supergravity by applying a series of transformations known as null Melvin twists \cite{gimon2003black}.\\

Thus, in order to describe gravity as a model of finite temperature condensed matter systems, we need to generalize as much as possible the AdS background. For instance, in condensed matter systems, a quantum critical point can be characterized by different types of critical exponents that satisfy different relationships between them. One example of these relations are the so-called \textit {hyperscaling relationships} where the dimension of space explicitly appears \cite{sachdev2007quantum}. In the last years there has been a great interest in studying non-relativistic systems that present this kind of property. As an illustration,  holographic properties of non-relativistic systems with hyperscaling violating symmetry and general dynamical exponent $ z $ are presented in \cite{kim2012schrodinger}, focusing the study on the entanglement entropy for Schrödinger-type backgrounds. \\

In general, zero-temperature systems with hyperscaling violating Schrödinger symmetry can be modeled by gravity duals using the following family of metrics \footnote{These backgrounds reduce to a pure Schrödinger geometry when $ \theta = 0 $.}
\begin{equation}
\label{hiper}
 ds^2_{D+3}=r^{-2\theta/D}\left[-b^2r^{2z}(dx^+)^2+\frac{1}{r^2}dr^2+r^2\big(dx^jdx_j+2dx^+dx^-\big)\right],\quad j=1,2,3...D.,
\end{equation} where $\theta$ is the \textit{hyperscaling violating exponent}.
This metric is not scale invariant, but transforms as
\begin{equation*}
\label{distancia}
ds \rightarrow \lambda^{\frac{\theta}{D}}ds.
\end{equation*}
There are several gravitational solutions that implement hyperscaling violating symmetry. For instance, a family of hyperscaling violating Lifshitz black holes in an Einstein-Maxwell-Dilaton theory are obtained in \cite{pedraza2019hyperscaling,alishahiha2012charged}, while Lifshitz black holes coupled to non-Abelian gauge fields with hyperscaling violating symmetry are studied in \cite{feng2015non}. Additionally, various aspects related to holographic theories with dynamical exponent and hyperscaling violating parameter are analyzed in \cite {dong2012aspects}.
\\

In the present manuscript we are interested in obtaining novel asymptotically Schrödinger black hole solutions. In order to achieve this aim we explore an Einstein-Maxwell-scalar setup. This theory has proven to support several black hole configurations with different asymptotic geometries. To give some examples, asymptotically AdS and Lifshitz black hole solutions within Einstein-Maxwell-scalar theories with different coupling functions and self-interaction scalar potentials are presented in \cite{yu2021constructing} and \cite{herrera2020black}, respectively. Meanwhile spherical black holes in Einstein-Maxwell-scalar models where the scalar field is coupled to the Maxwell invariant are discussed in \cite{astefanesei2019einstein}. Additionally, a rotating configuration that generalizes Lifshitz black holes is presented within the framework of the Einstein-Maxwell-Dilaton theory in \cite{herrera2021rotating}. Furthermore, here we show that within a coupled Maxwell-scalar fields setup it is possible to construct asymptotically Schrödinger  black hole backgrounds with hyperscaling violating symmetry for an arbitrary dynamical exponent $z$.
\\

\textit{Organization of the paper}. In Sec. \ref{hv} we introduce our gravitational framework and obtain the field equations in a generic hyperscaling violating Schrödinger background. We further construct a new family of exact hyperscaling violating black holes by fixing the self-interaction scalar potential in Subsec \ref{bh2}. In order to elucidate under which conditions our solutions are asymptotically regular, in Subsec. \ref{invar} an asymptotic analysis of the curvature invariants is presented. We conclude with some remarks and a brief discussion in Sec. \ref{con}.

\section{Hyperscaling violating asymptotically Schrödinger black holes }
\label{hv}

In this section we introduce our gravity  model and construct hyperscaling violating Schrödinger black holes for this theory. Our starting point is an Einstein-Maxwell-scalar theory described by the following gravitational action in $ D + 3 $ dimensions (we set natural units)
\begin{equation}
 \label{accion1}
S[g_{\mu\nu},A_\mu,\phi]=\frac{1}{16 \pi G}\int d^{D+3}x\sqrt{-g}\Bigg[R-\frac{1}{2}(\nabla \phi)^2-V(\phi)-\frac{1}{4}\chi_i(\phi)F_i^2\Bigg],
\end{equation}
where $R$ is the Ricci scalar.  This action minimally couples gravity to the matter fields, namely, to a real scalar field $\phi$ with a self-interaction potential $V(\phi)$, and to $n$ Maxwell fields $F_i$ associated to the  vector potentials $A_i$
\begin{align*}
F_i= dA_i\rightarrow \quad F_{\mu \nu}^{i}=\partial_\mu A_\nu^i - \partial_\nu A_\mu^i , \quad i=1,2,3 \ldots n.
\end{align*}
The  functions $\chi_i(\phi)$ govern the coupling of $\phi$ to the electromagnetic fields $F_{\mu \nu}^{i}$. The scalar field self-interaction potential together with the coupling functions are crucial ingredients for obtaining analytic black hole configurations within our theory. The field equations obtained by varying the above action with respect to the field variables $\phi$, $A_\mu$ and $g_{\mu\nu}$ are
\begin{equation}
\Box \phi=\partial_\phi V(\phi)+\frac{1}{4}\partial_\phi\chi_i(\phi)F^{2}_i,\quad					
\nabla_\mu\left[\chi_{(i)}(\phi)F_{(i)}^{\mu \nu}\right]=0,\quad
\label{proca25}
  \mathcal{E}_{\mu\nu}=: R_{\mu \nu}-\frac{1}{2}R g_{\mu\nu}-T_{\mu\nu}=0,
\end{equation}
where the energy-momentum tensor of the matter fields acting as the source of the curvature of spacetime is
\begin{equation}
 T_{\mu\nu}=-\frac{1}{4}g_{\mu\nu}{(\nabla\phi)}^2-\frac{1}{2}g_{\mu\nu}V(\phi)+\frac{1}{2}\chi_i(\phi)\left[F^{(i)}_{\alpha\mu}F^{\alpha}_{(i)\nu}-\frac{1}{4}F^2_ig_{\mu\nu}\right]+\frac{1}{2}\partial_\mu\phi\partial_\nu\phi.
 \label{energiamomento25}
\end{equation}
 
To construct exact black hole solutions, it is convenient to consider the following stationary Schrödinger type \textit{ansatz} with a single blackening factor $ f (r) $ and with hyperscaling violating symmetry
\begin{equation}
\label{metrica2}
    ds^2_{D+3}=r^{-2\tilde{\theta}/D}\left[-b^2r^{2z}f(r)(dx^+)^2+\frac{1}{r^2f(r)}dr^2+r^2\big(dx^jdx_j+2dx^+dx^-\big)\right],\quad j=1,2,3...D.
\end{equation}Thus, the metric (\ref{metrica2}) is asymptotically Schrödinger as long as $ f (r) \rightarrow 1 $ near the asymptotic boundary. It is worth mentioning that (\ref{metrica2}) describes a stationary spacetime, so by consistency, it is imposed that all matter fields inherit the isometries of the spacetime, i.e. the $n$ vector potentials and the scalar field are only functions of the holographic coordinate $r$. Furthermore the vector potential \textit{ansätze} compatible with the symmetries of the (\ref{metrica2}) contain purely electric terms, i.e,
\begin{equation*}
\phi\equiv\phi(r),\quad\quad A_i\equiv A_i(r)dx^+.
\end{equation*}
Plugging it into the field equations (\ref{proca25}) it follows that the Maxwell equations accept a trivial first integral
\begin{equation}
{ A_i}^\prime(r)=\frac{\rho_i}{\sqrt{f}\chi_i(\phi)r^{(1+\theta)+D(1-\theta)}},\label{p22222222}
\end{equation}
where $ \rho_i $ are arbitrary integration constants that are associated to the electric charges of the Maxwell fields. Henceforth a prime denotes a derivative with respect to $r$.
The non-trivial Einstein field equations read
\begin{equation}
\mathcal{E}^-_+:\quad r^2f^{\prime\prime}+\frac{1}{2}{\frac{r^2f^{\prime2}}{f}}+\bigg\{5z-\theta(D+1)+(D-2)\bigg\}rf^\prime+\bigg\{2(z-1)[D+2z-\theta(D+1)]\bigg\}f=\frac{\rho_i^2r^{2\left[D(\theta-1)-z\right]}}{fb^2\chi_i(\phi)},\label{u1111}\end{equation} 
\begin{equation}
 \mathcal{E}^r_r:\quad (D+1)(D+2)(\theta-1)^2f=-\frac{1}{r^{2\theta}}\left[\frac{{-(\nabla\phi)}^2}{2}+V(\phi)\right],\label{u2222}\end{equation}  
\begin{equation}
\mathcal{E}^+_+:\quad(\theta-1)(D+1)rf^\prime+(\theta-1)(D+1)\left\{D(1-\theta)+2\right\}f=\frac{1}{r^{2\theta}}\left[\frac{{(\nabla\phi)}^2}{2}+V(\phi)\right],\label{u3333}\end{equation}
where the hyperscaling violating exponent is normalized with the number of transversal spatial dimensions, i.e. $\theta\equiv\tilde{\theta}/D$. Moreover, the scalar field equation reads \begin{equation}
\phi^{\prime\prime}+\left\{\frac{f^\prime}{2f}+\frac{(D+3)-\theta(D+1)}{r}\right\rbrace\phi^\prime=\frac{\partial_\phi V(\phi)}{fr^{2(1+\theta)}},\label{scalar}
\end{equation}
Notice that as a result of our \textit{ansätze} for the vector potentials\footnote{With purely electric vector potentials the Maxwell invariants $F_i^2$ are identically zero, this is a direct consequence of our spacetime symmetries.}, the coupling functions $\chi_i(\phi)$ do not appear in (\ref{scalar}). Hence, our coupling functions will not be determined by the field equations, so in principle they are completely arbitrary. This freedom is crucial in order to find black holes solutions. 
To decouple the scalar field and its self-interaction potential, the following linear combinations are taken
\begin{equation}
\mathcal{E}^r_{r}+\mathcal{E}^+_{+}:\quad  
(\phi^\prime)^2=\frac{(D+1)(\theta-1)}{r^2}\bigg[\frac{rf^\prime}{f}+2\theta\bigg],\label{n1}
\end{equation}
\begin{equation}
\mathcal{E}^r_{r}-\mathcal{E}^+_{+}:\quad  
V(\phi)=\frac{(D+1)(\theta-1)}{2}r^{2\theta}\Bigg\{rf^\prime-2\left[\theta(D+1)-(D+2)\right]f\Bigg\}.\label{v1}
\end{equation}
From the field equations we notice that the metric function $f(r)$ cannot be decoupled from the matter fields, this is a substantial difference comparing to configurations with Lifshitz symmetry \cite{herrera2020black, pedraza2019hyperscaling, alishahiha2012charged}. Thus, the scalar field and its self-interaction potential are determined by the metric function and its first derivative. Furthermore, the Maxwell fields are determined by the square root of the metric function and the coupling functions. 

\subsection{Black hole solutions 
}
\label{bh2}
In order to obtain an exact black hole configuration, we propose the following \textit{ansatz} for the scalar field self-interaction potential
\begin{equation}
    V(\phi)=V_o r^{2\theta},\label{selfi}
\end{equation}
where $ V_o $ is an arbitrary constant. Hence, this \textit{ansatz} allows us to decouple the metric function from (\ref{v1}). Thereby, by inserting (\ref{selfi}) into (\ref{v1}) we obtain the following equation
\begin{equation}
   rf^\prime-2\left[\theta(D+1)-(D+2)\right]f -\frac{2V_o}{(D+1)(\theta-1)}=0,\label{v22}
\end{equation}
which is a first order inhomogeneous Euler differential equation whose general solution is
\begin{equation}
    f(r)=cr^{2\left[\theta(D+1)-(D+2)\right]}+\frac{V_o}{(D+1)(\theta-1)\left[D-\theta(D+1)+2\right]},\label{v3}
\end{equation}
where $c$ is a generic integration constant. From (\ref{v3}) one can observe that in order to be able to construct this solution it is crucial to have a non-trivial self-interaction potential.
In order for our metric function to be well-defined, the following restrictions for the parameters of the solution must be met 
 \begin{equation}
   \theta\not=1, \qquad  \qquad   \theta\not=\frac{D+2}{D+1}.
 \end{equation}

Furthermore, it is necessary to ensure the appropriate asymptotic limit of the metric (\ref{metrica2}). This imposes a constraint on the constant $ V_o $
\begin{equation}
   V_o=(D+1)(\theta-1)\left[D-\theta(D+1)+2\right].\label{const}
\end{equation}
Notice that when $ \theta = 0 $ the asymptotic constraint (\ref{const}) reduces $ V_o $ to the typical value of a negative cosmological constant for a Schrödinger spacetime \cite{son2008toward}. However, this limit leads to a complex scalar field as it can be seen from (\ref{n1}).

Thus, by setting the integration constant $c$ to the event horizon position, the metric function $f$ takes the following form
\begin{equation}
    f(r)=1-\bigg(\frac{r_h}{r}\bigg)^{2\left[(D+2)-\theta(D+1)\right]}.\label{hn3}
\end{equation}
This solution represents a gravitational configuration depending on both the hyperscaling violating exponent $\theta$ and an arbitrary number of spatial transversal dimensions. Moreover, notice that (\ref{hn3}) does not depend on the dynamical exponent $z$. This is a direct consequence of the fact that in the Ricci tensor $R ^ \mu_\nu $ of Schr\"odinger-type spacetimes, the dynamical critical exponent only appears in its $ {R} ^ -_ {+} $ component  \cite{kim2012schrodinger}. 

Now we proceed to compute the scalar field expression by inserting (\ref{hn3}) into (\ref{n1}) and integrating with respect to $ r $. As it was previously emphasized, in our field equations $\phi'$ depends only on the metric function and its first derivative. Thereby, we expect a more complicated solution compared to the results that have been reported in Einstein-Maxwell-Dilaton Lifshitz black hole type  solutions \cite{herrera2020black,alishahiha2012charged,herrera2021rotating,pedraza2019hyperscaling}. Concretely, we find  
 \begin{eqnarray}
\phi(r)&=&\gamma\Bigg\{\log\left[\frac{2-(1+\beta)\bigg(\frac{r_h}{r}\bigg)^{2\left[(D+2)-\theta(D+1)\right]}+2W(r)}{\bigg(\frac{r_h}{r}\bigg)^{2\left[(D+2)-\theta(D+1)\right]}}\right]
\\
&+&\sqrt{\beta}\log\left[1+\beta-2\beta\bigg(\frac{r_h}{r}\bigg)^{2\left[(D+2)-\theta(D+1)\right]}-2\sqrt{\beta}W(r)\right]\Bigg\},\label{kop3}
\nonumber
\end{eqnarray}
where the following function has been defined
\begin{equation*}
W(r)=\sqrt{1-\beta\bigg(\frac{r_h}{r}\bigg)^{2\left[(D+2)-\theta(D+1)\right]}}\sqrt{1-\bigg(\frac{r_h}{r}\bigg)^{2\left[(D+2)-\theta(D+1)\right]}}\quad  
\end{equation*}
with the following constants that depend only on $\theta$ and $D$
\begin{equation*}
    \beta=\frac{(D+2)(\theta-1)}{\theta}, \quad\gamma=\frac{\sqrt{2\theta(D+1)(\theta-1)}}{2\left[(D+2)-\theta(D+1)\right]}. 
\end{equation*}
 The scalar field profile (\ref{kop3}) does not depend on the dynamical exponent $z$ and cannot be inverted in the form $ r(\phi) $. Thus, it is not possible to explicitly express the coupling function $ \chi (\phi) $ nor the potential $ V (\phi) $ as functions of the scalar field. 
 Furthermore, one must note that $\phi$ diverges when $ \theta \rightarrow 0 $. Consequently, the hyperscaling violating parameter is crucial for the existence of a dynamical scalar field in our spacetime. Moreover, we must impose\footnote{When $ \beta = 0 $  the scalar field only has a single logarithmic branch.} the conditions $ \beta \geq 0 $ and $\theta(D+1)(\theta-1)>0$ in order to preserve the real character of the scalar field. The latter restriction implies that $ \theta <0 $ or $ \theta> 1 $. More importantly, according to the admisible values of $ \beta $, the scalar field $ \phi $ possesses a well-defined dynamics in different sectors of the spacetime. This fact is summarized in Table \ref{tab:my_label4}.
   
 \begin{table}[H]
\centering
\begin{tabular}{|c|c|c|c|c|} 
\hline
parameter restrictions                            & $\beta$ value &spacetime sector where $ \phi $ is well-defined   \\ \hline
                                                  &                &                                             \\
$\theta<0$                                        & $\beta>1$     & $r> r_h$                                    \\
                                                  &                &                                             \\
$0<\theta<1$                     & $\beta<0$     & not defined                                 \\
                                                  &                &                                             \\
$1\leq\theta<\frac{D+2}{D+1}$ &$0\leq\beta<1$ & $r>r_h$                                     \\
                                                  &                &                                             \\
$\theta>\frac{D+2}{D+1}$                       & $\beta>1$     & $r< r_h$                                    \\ 
                                                  &                &                                             \\ \hline
\end{tabular}
\caption{Scalar field dynamics for our black hole family of solutions.}
\label{tab:my_label4}
\end{table}
One must ensure that the scalar field dynamics is always compatible with the spacetime sector where the blackening factor (\ref{hn3}) preserves the metric signature (\ref{metrica2}). Thereby, in principle all restrictions of Table \ref{tab:my_label4} are valid as well for the metric function $f$. \\

Once the scalar field dynamics supported within our hyperscaling violating spacetime is elucidated, it remains to determine the coupling functions of the theory. These functions  are restricted by the equation  (\ref{u1111}), being the simplest case the one with a scalar field coupled to a single Maxwell field. Consequently, by using (\ref{u1111}) the coupling function is implicitly determined as
  
\begin{equation}
    \chi(\phi)=\frac{\delta r^{2\left[D(\theta-1)-z\right]}}{r^2ff^{\prime\prime}+\frac{1}{2}{(rf^{\prime})^2}+\bigg[5z-\theta(D+1)+(D-2)\bigg]rff^\prime+\bigg[2(z-1)[D+2z-\theta(D+1)]\bigg]f^2},\label{fghs}
\end{equation}
where $\delta=\frac{\rho^2}{b^2}$. For the solution (\ref{hn3}) this coupling function adopts the following form
\begin{equation}
    \chi(\phi)=\frac{\tilde{\delta}r^{2\left[D(\theta-1)-z\right]}}{a_0+a_1\bigg(\frac{r_h}{r}\bigg)^{2\left[(D+2)-\theta(D+1)\right]}+a_2\bigg(\frac{r_h}{r}\bigg)^{4\left[(D+2)-\theta(D+1)\right]}},\label{copl}
\end{equation}
where we have introduced the following constants
\begin{align*}
  a_0&=2(z-1)\left[D+2z-\theta(D+1)\right],
  \\a_2&=4\left[-3+z+D(\theta-1)+\theta\right]^2,
\end{align*}
\begin{equation*}
 a_1=2\bigg[-14-7D-D^2+14z+3Dz-4z^2+\theta(D+1)(7+2D-3z)\\-{\theta}^2(D+1)^2\bigg].
\end{equation*}
 In addition, both the scalar field self-interaction potential (\ref{selfi}) and the coupling function (\ref{copl}) reduce to a finite constant in the event horizon. Finally, it remains to determine the vector potential of the theory. This quantity is calculated by inserting the coupling function (\ref{copl}) and the metric function (\ref{hn3}) into Maxwell's equations (\ref{p22222222}) and integrating the Maxwell field with respect to  $ r $. We find the following vector potential
\begin{equation}
    A(r)=Q r^{2z-\theta(D+1)+D}\sqrt{1-\bigg(\frac{r_h}{r}\bigg)^{2\left[(D+2)-\theta(D+1)\right]}}\Bigg\{(z-1)-\left[z-3+\theta(D+1)-D\right]\bigg(\frac{r_h}{r}\bigg)^{2\left[(D+2)-\theta(D+1)\right]}\Bigg\}
    \label{m2}
\end{equation}
with the following constant
    $Q=\frac{b^2}{\rho}$.

Summarizing, we have constructed a novel family of hyperscaling violating black holes with Schrödinger symmetry. This configuration corresponds to an exact solution of the field equations (\ref{p22222222})-(\ref{scalar}). The solution consists of a gravitational metric (\ref{metrica2}) with a blackening factor (\ref{hn3}), a real scalar field (\ref{kop3}) with a self-interaction potential (\ref{selfi}) which is coupled to a Maxwell field generated by the vector potential (\ref{m2}) through the coupling function (\ref{copl}).

\subsection{Curvature invariants}
\label{invar}

We now analyze the curvature invariants of the spacetime described by the metric (\ref{metrica2}) in order to investigate under which circumstances our field configuration represents asymptotically regular black holes.

\begin{figure}[H]
\centering
\subfloat[]{%
  \includegraphics[clip,width=0.5\columnwidth]{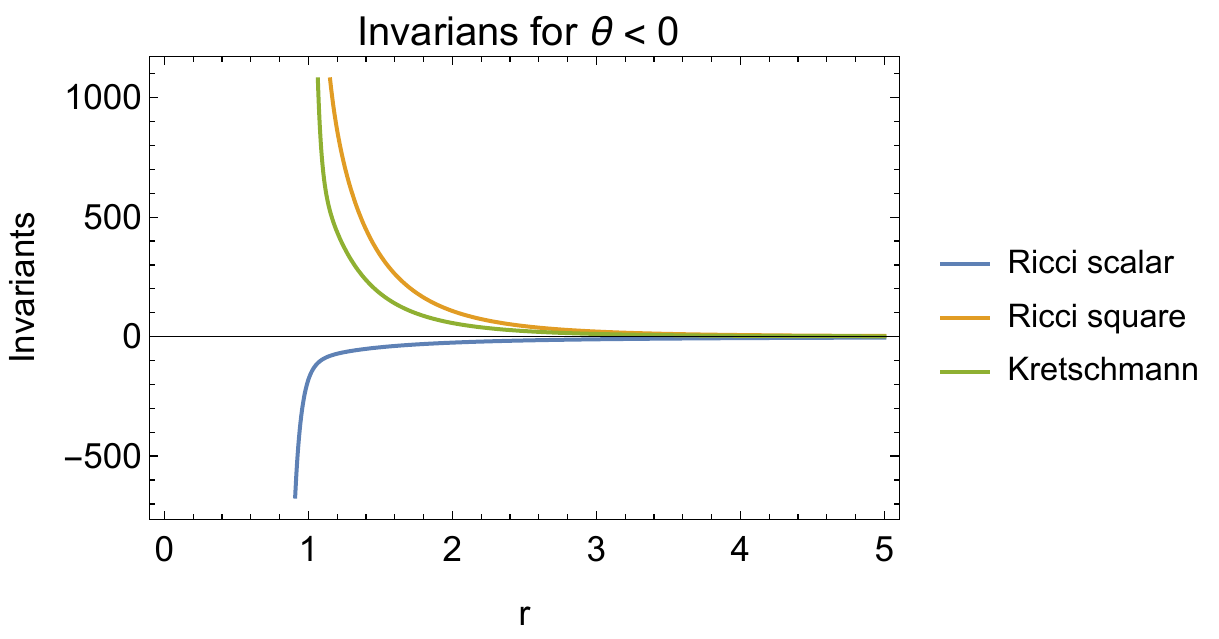}%
}
\subfloat[]{%
  \includegraphics[clip,width=0.5\columnwidth]{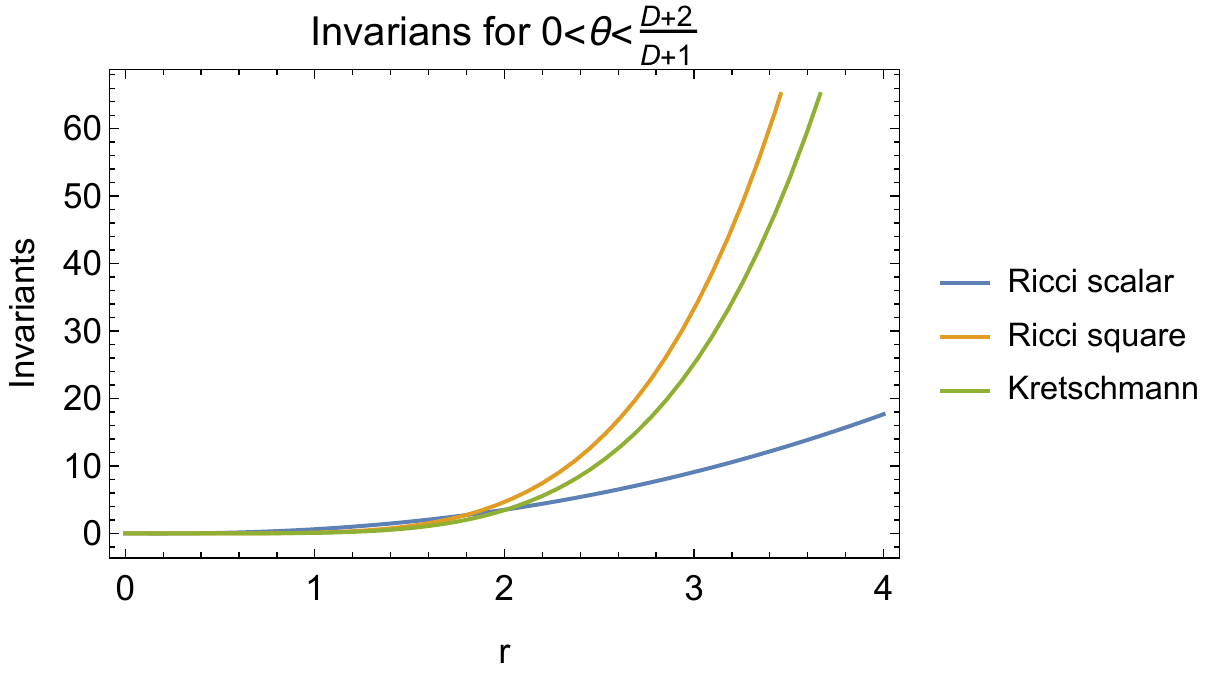}%
}\\
\subfloat[]{%
  \includegraphics[clip,width=0.5\columnwidth]{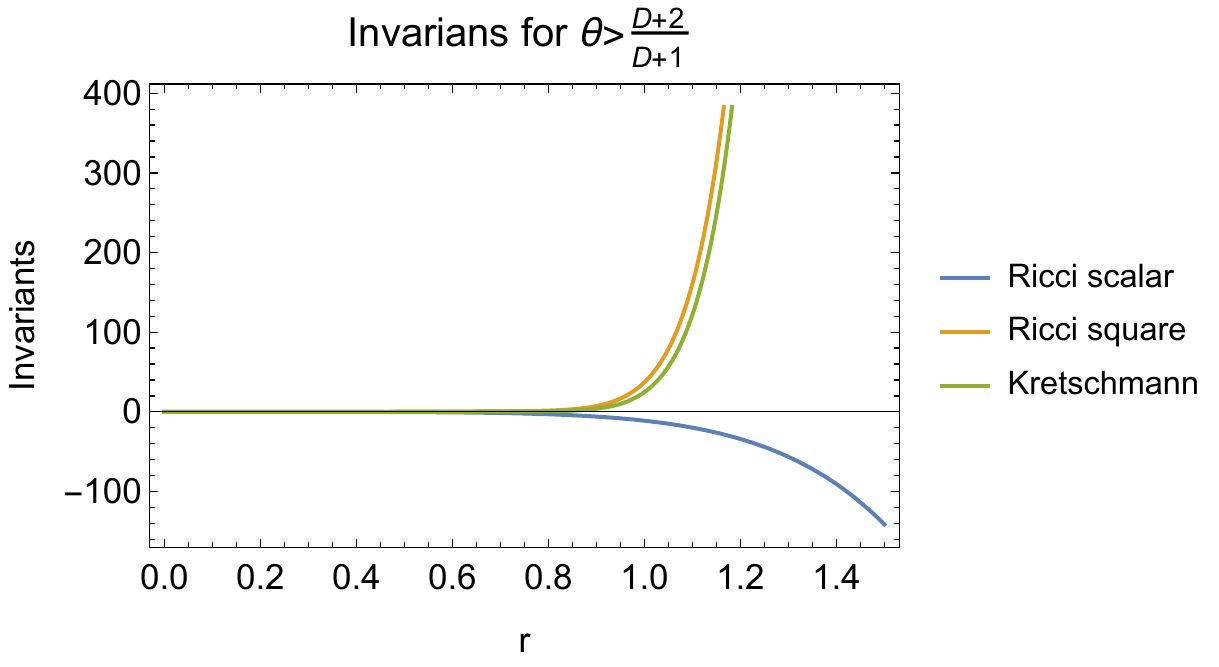}%
}
\subfloat[]{%
  \includegraphics[clip,width=0.5\columnwidth]{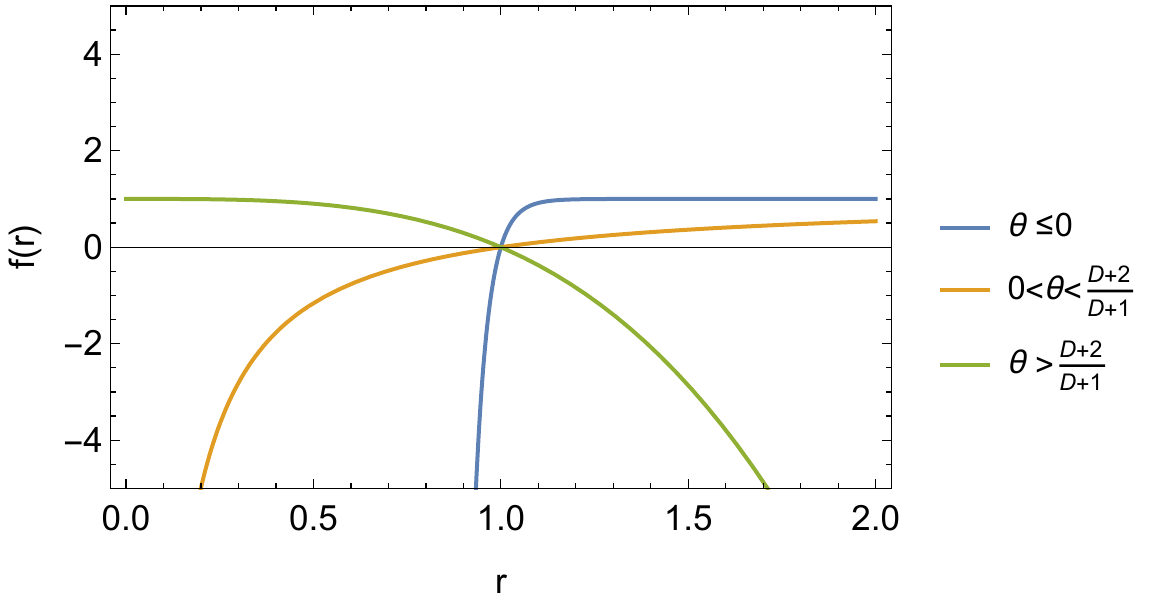}
  }%
\caption{The curvature invariants and metric function for our black hole family. We set $r_h=1$ and $D=3$. (a) $\theta\leq 0$ case, (b) $\theta<\frac{D+2}{D+1}$ case, (c) $\theta>\frac{D+2}{D+1}$ case, (d) The metric function for all the aforementioned ranges.}
\label{AFM1}
\end{figure}

The curvature invariants are a set of scalar quantities constructed from the Riemann tensor, Ricci tensor or the Weyl tensor and allow us to study the geometry in a deeper way since they are independent of the coordinate system quantities 
(see for instance the curvature invariants for hyperscaling violating Lifshitz black holes \cite {pedraza2019hyperscaling}). Particularly, when black holes exist in our geometry,  it is crucial to identify the physical singularities of the spacetime. 
Here we realize an analysis of the Ricci scalar $R$, the square of the Ricci tensor $ \mathfrak {R}: = R_{\mu\nu} R ^ {\mu\nu} $ and the Kretschmann invariant $ \mathcal {K}: = R_{abcd} R ^{abcd} $ to identify genuine physical singularities of the black hole configurations constructed in \ref{bh2}. 
For our metric (\ref{metrica2}) the curvature invariants read

   \begin{equation}
   R=(D+2)(\theta-1)r^{2\theta}\Bigg\{rf'+\left[2-(\theta-1)(D+1)\right]f \Bigg\},
\label{invth1}
\end{equation}\\ 
\begin{equation}
\mathfrak {R}=(D+2)(\theta-1)^2r^{4\theta}\Bigg\{\kappa_1f^2+rf'\bigg[\kappa_2f+\frac{D+3}{4}rf'\bigg]\Bigg\},\label{ksm2}
\end{equation}\\
\begin{equation}
 \mathcal{K}=(\theta-1)^2r^{4\theta}\Bigg\{\kappa_3f^2+(D+2)rf'\bigg[4f+rf'\bigg]\Bigg\},\label{ksm11}
\end{equation}
with the following constans that depend only on $\theta$ and $D$
\begin{align*}
\kappa_1&=(D+2)+\left[\theta(D+1)-(D+2)\right]^2,\\ \kappa_2&=(D+3)-(D+1)(\theta-1),\\
\kappa_3&=-2D^2(\theta-1)^2+D^3(\theta-1)^2+2\left[3+\theta(\theta-2)\right]+D\left[7+5\theta(\theta-2)\right].
\end{align*}
From the above expressions it is evident that either by adding the hyperscaling violating parameter or by considering black holes within the (\ref{metrica2}) spacetime the curvature invariants are no longer constant as it was the case for purely Lifshitz and Schrödinger spacetimes \cite {dong2012aspects, kim2012schrodinger}. The expressions (\ref{invth1})-(\ref{ksm11}) for our solution (\ref{hn3}) read
 
 \begin{equation}
  R=(D+2)(\theta-1)r^{2\theta}\Bigg\{\omega_0+\omega_1\bigg(\frac{r_h}{r}\bigg)^{2\left[(D+2)-\theta(D+1)\right]}\Bigg\} ,\label{ikl111}
\end{equation}\\
\begin{equation}
\mathfrak {R}=(D+2)(\theta-1)^2r^{4\theta}\Bigg\{\omega_2+\omega_3\bigg(\frac{r_h}{r}\bigg)^{2\left[(D+2)-\theta(D+1)\right]}+\omega_4\bigg(\frac{r_h}{r}\bigg)^{4\left[(D+2)-\theta(D+1)\right]}\Bigg\},\label{ikl222} 
\end{equation}\\
\begin{equation}
\mathcal{K}=2(\theta-1)^2r^{4\theta}\Bigg\{\omega_5+\omega_6\bigg(\frac{r_h}{r}\bigg)^{2\left[(D+2)-\theta(D+1)\right]}+\omega_7\bigg(\frac{r_h}{r}\bigg)^{4\left[(D+2)-\theta(D+1)\right]}\Bigg\}.\label{ikl333}
\end{equation}
 with the following constants that depend on $\theta$ and $D$
 \begin{equation*}
     \omega_0=(D+3)-\theta(D+1),\quad\omega_1=(D+1)(1-\theta),\quad\omega_2=D^2+5D+6-2\theta(D+1)(D+2)+\theta^2(D+1)^2,
 \end{equation*}
 \begin{equation*}
     \omega_3=-2(\theta-1)(D+1)(D+2),\quad\omega_4=(\theta-1)^2(D+1)^2(D+2),
 \end{equation*}
 \begin{equation*}
     \omega_5=D^3-2D^2+7D+6-\theta(2D^3-4D^2+10D+4)+\theta^2(D^3-2D^2+5D+2),
 \end{equation*}
 \begin{equation*}
     \omega_6=-2(\theta-1)\left\{2(\theta+1)+D\left[1+D(4+D(\theta-1)-2\theta))+5\theta\right]\right\},
 \end{equation*}
 \begin{equation*}
    \omega_7=3(\theta-1)^2\left\{2+D\left[5+D(D+2)\right]\right\}.
\end{equation*}
Thus, for our family of black holes the curvature invariants do not depend on  the dynamical exponent $z$ since the metric function is independent of this parameter. In order to identify the asymptotic boundary of the manifold and the physical singularity of our spacetime, in Fig. \ref{AFM1} the curvature invariants and the metric function are plotted for different ranges of the parameter $\theta$. Without loss of generality we set the horizon position to $r_h=1$ and the number of spatial transversal dimensions to $D=3$. However, the same asymptotic behavior is observed for different values of $D$.

\vspace{0.5in}
From Fig. \ref{AFM1} one observes the asymptotic behaviour summarized in Table \ref{bn1}.
 \begin{table}[H]
     \centering
     \begin{tabular}{|c|c|c|c|c|c|} 
     \hline
     parameter restrictions& spacetime singularity &spacetime boundary & metric function on the boundary   \\
     \hline
 $\theta\leq0$ & $r\rightarrow0$ &$r\rightarrow\infty$ &$f(r)\rightarrow1$  \\
 &     &             &\\
 $0<\theta<\frac{D+2}{D+1}$ &$r\rightarrow\infty$&$r\rightarrow0$& $f(r)\rightarrow-\infty$ \\
  &     &             &\\
 $\theta>\frac{D+2}{D+1}$ & $r\rightarrow\infty$ &$r\rightarrow0$& $f(r)\rightarrow1$ \\  \hline
 
     \end{tabular}
     \caption{Asymptotic analysis of the spacetime curvature invariants for our family of black holes.}
     \label{bn1}
 \end{table}
According to the results presented in Table \ref{bn1} we argue that our configuration \ref{bh2} is asymptotically regular only for $ \theta \leq0 $ or $ \theta> \frac {D + 2} {D + 1} $. For these cases, the singularity of the spacetime is hidden behind an event horizon and the blackening factor (\ref{hn3}) preserves the metric signature (\ref{metrica2}).

Finally, by comparing the parameter restrictions due to the asymptotical analysis of the curvature invariants to the constrains obtained from the scalar field dynamics (summarized in Table \ref{tab:my_label4}), we conclude that the allowed physical configurations are restricted to $ \theta <0 $ or $ \theta> \frac {D + 2} {D + 1} $. In view of this fact, the spacetime is not asymptotically regular within the range $0\le\theta\le\frac{D+2}{D+1}$.

 \section{Conclusions and final remarks}
\label{con}

We have constructed a new family of exact hyperscaling violating black holes with Schrödinger symmetry within an Einstein-Maxwell-scalar theory. According to our understanding, in this paper we report the first gravitational asymptotically Schrödinger black hole configuration for any dinamical critical exponent $z$ in an arbitrary number of spacetime dimensions. Nevertheless, an analysis of the curvature invariants revealed that the values of the hyperscaling parameter $\theta$ must be constrained for our solution to be asymptotically regular. In this work we illustrate how a self-interaction potential for the scalar field and at least one non-minimally coupled Maxwell field are needed in order to support exact black hole solutions within considered setup. \\

The solution we have constructed here represents a gravitational background which is dual to non-relativistic field theories with hyperscaling violating Schrödinger symmetry at finite temperature within the framework of the Gravity/Condensed Matter Theory correspondence .
\\

There are many research directions that can be followed as direct extensions of this work. The most immediate one would be to generalize the solution constructed here by including a second Maxwell field that allows to obtain a second term with an extra integration constant in the metric function. In addition, it will be interesting to use a  more generic metric $ansatz$ that includes rotation parameters, so that we would be able to construct configurations with several angular momenta in the spirit of \cite{herrera2021rotating}. Moreover, a complete thermodynamic analysis of our solution is needed in order to see is our black hole configurations are thermodynamically consistent. For this purpose we can exploit the so-called generalized ADT quasi-local method, foremost introduced in \cite{kq}. This method has been implemented and proved to be well suited in the construction of conserved charges in higher-order gravity theories (see for instance \cite{a1,a2,a3,a4}), as well as in Einstein-Maxwell-Dilaton theories \cite{a5,herrera2020black,herrera2021rotating}.\vspace{0.8in}

{\it Acknowledgements.} The authors would like to thank Manuel de la Cruz López, Jhony A. Herrera-Mendoza, Daniel F. Higuita-Borja, Julio A. Méndez-Zavaleta and Uriel Noriega Cornelio for fruitful discussions. A.H.-A. thanks the Sistema Nacional de Investigadores (SNI) for support. The authors acknowledge financial support from a CONACYT Grant No. A1-S-38041. C.E. R.-F. also acknowledges support from a CONACYT Grant No. 743421.

\bibliographystyle{unsrt}
\bibliography{referencias3.bib}

\end{document}